\documentclass{iau} 
\usepackage{graphicx}

\title[The 2X-HI disks of spiral galaxies]
      {The 2X-H\,{\sc i} disks of spiral galaxies}

\author[B\"arbel S. Koribalski] 
       {B\"arbel S. Koribalski}

\affiliation{Australia Telescope National Facility, CSIRO
  Astronomy \& Space Science \\ P.O. Box 76, Epping, NSW 1710, Australia \\
  email: {\tt Baerbel.Koribalski@csiro.au}}

\pubyear{2016}
\volume{321} 
\jname{Formation and Evolution of Galaxy Outskirts}
\editors{Armando Gil de Paz, Johan Knapen \& Janice Lee, eds.}
\begin{document}

\maketitle

\begin{abstract}
The outskirts of galaxies --- especially the very extended H\,{\sc i} disks 
of galaxies --- are strongly affected by their local environment. I highlight
the giant 2X-H\,{\sc i} disks of nearby galaxies (M\,83, NGC~3621, and 
NGC~1512), studied as part of the Local Volume H\,{\sc i} Survey (LVHIS), 
their kinematics and relation to XUV disks, signatures of tidal interactions 
and accretion events, the $M_{\rm HI} - D_{\rm HI}$ relation as well as the 
formation of tidal dwarf galaxies. - Using multi-wavelength data, I create 3D 
visualisations of the gas and stars in galaxies, with the shape of their 
warped disks obtained through kinematic modelling of their H\,{\sc i} velocity
fields.

\end{abstract}

\firstsection 

\section{Introduction}

To trace gas and stars in the outskirts of galaxy disks, we typically use
H\,{\sc i} mapping (e.g., \cite[Huchtmeier \& Bohnenstengel 1981]{HB1981};
\cite[Koribalski \& L\'opez-S\'anchez 2009]{KL2009}; \cite[Heald et al. 
2011]{Heald2011}; \cite[Serra et al. 2012]{Serra2012}; \cite[Lee-Waddell et 
al. 2012]{LW2012}; \cite[Koribalski et al. 2016]{K2016}) and deep, wide-field 
optical imaging (e.g., \cite[Mart\'inez-Delgado et al. 2010]{MD2010}; 
\cite[Duc et al. 2015]{Duc2015}). The H\,{\sc i} disks of galaxies, which 
often extend a factor two or more beyond the bright stellar disk 
(\cite[Warren et al. 2004]{WJK2004}), are excellent tracers of their total 
mass (visible and dark matter). The gas kinematics allows us to model their
3D shapes and obtain rotation curves, from which the radial distribution 
and amount of dark matter is derived.

GALEX ultra-violet (UV) imaging of nearby galaxies led to the discovery of 
extended UV disks (XUV-disks) in the galaxies M\,83 and NGC~4625 (\cite[Thilker
et al. 2005]{Thilker2005}; \cite[Gil de Paz et al. 2005]{Gil2005}), indicating 
star formation well beyond the radius where H\,{\sc ii} regions are typically 
found. Here I suggest that the H\,{\sc i} distributions of XUV-disk galaxies 
extend about twice as far. The term 2X-H\,{\sc i} disk was introduced by 
\cite{KL2009}, who found the giant H\,{\sc i} disk of NGC~1512 to extend a 
factor two beyond its XUV-disk. Another prominent spiral galaxy with an 
2X-H\,{\sc i} disk is M\,83 (\cite[Koribalski 2008]{K2008}; \cite[Jarrett et 
al. 2013]{J2013}; \cite[Koribalski 2015]{K2015}).

\begin{figure}
\begin{tabular}{cc}
 \includegraphics[width=7.9cm]{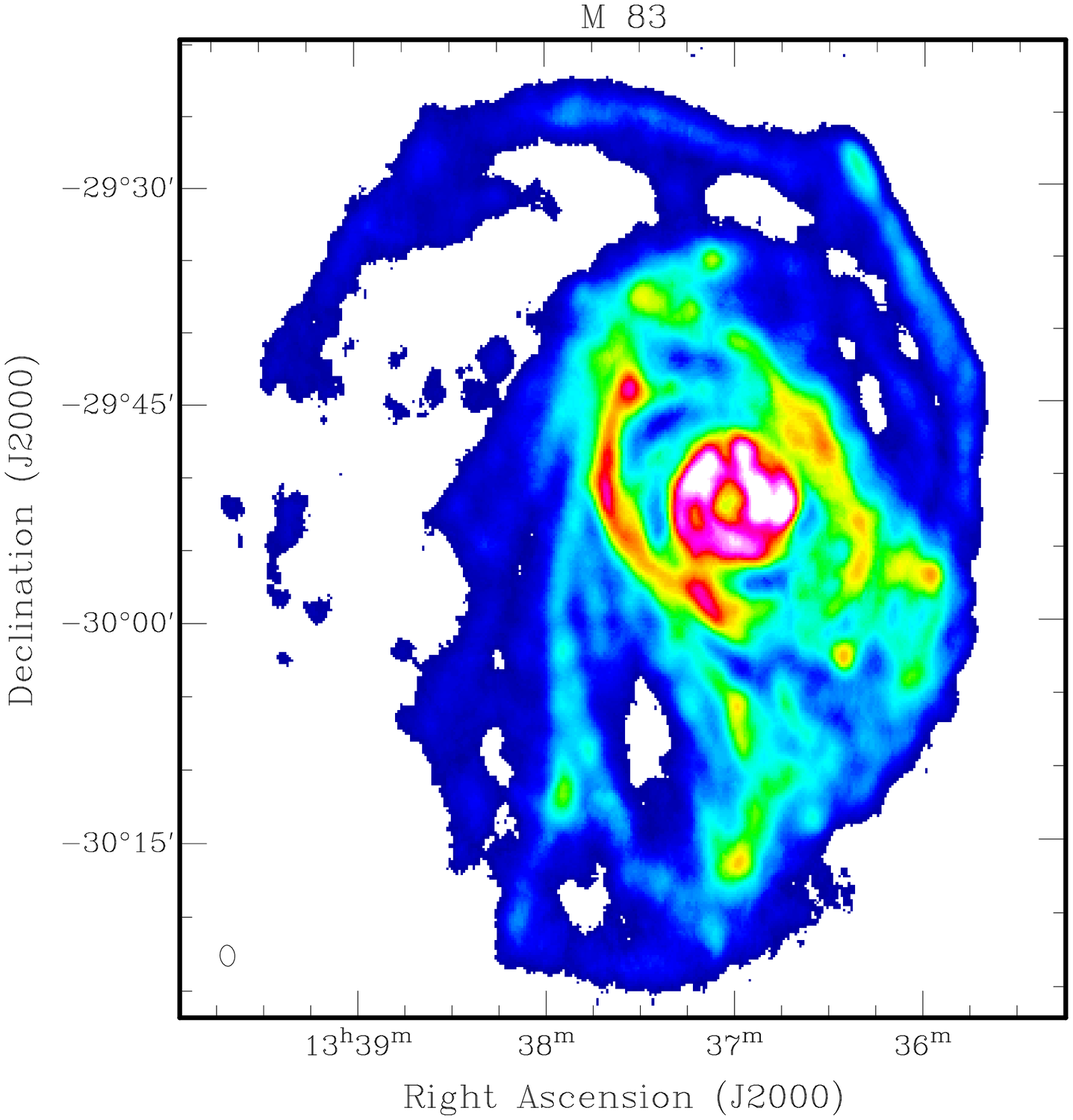} &
 \includegraphics[width=5.5cm]{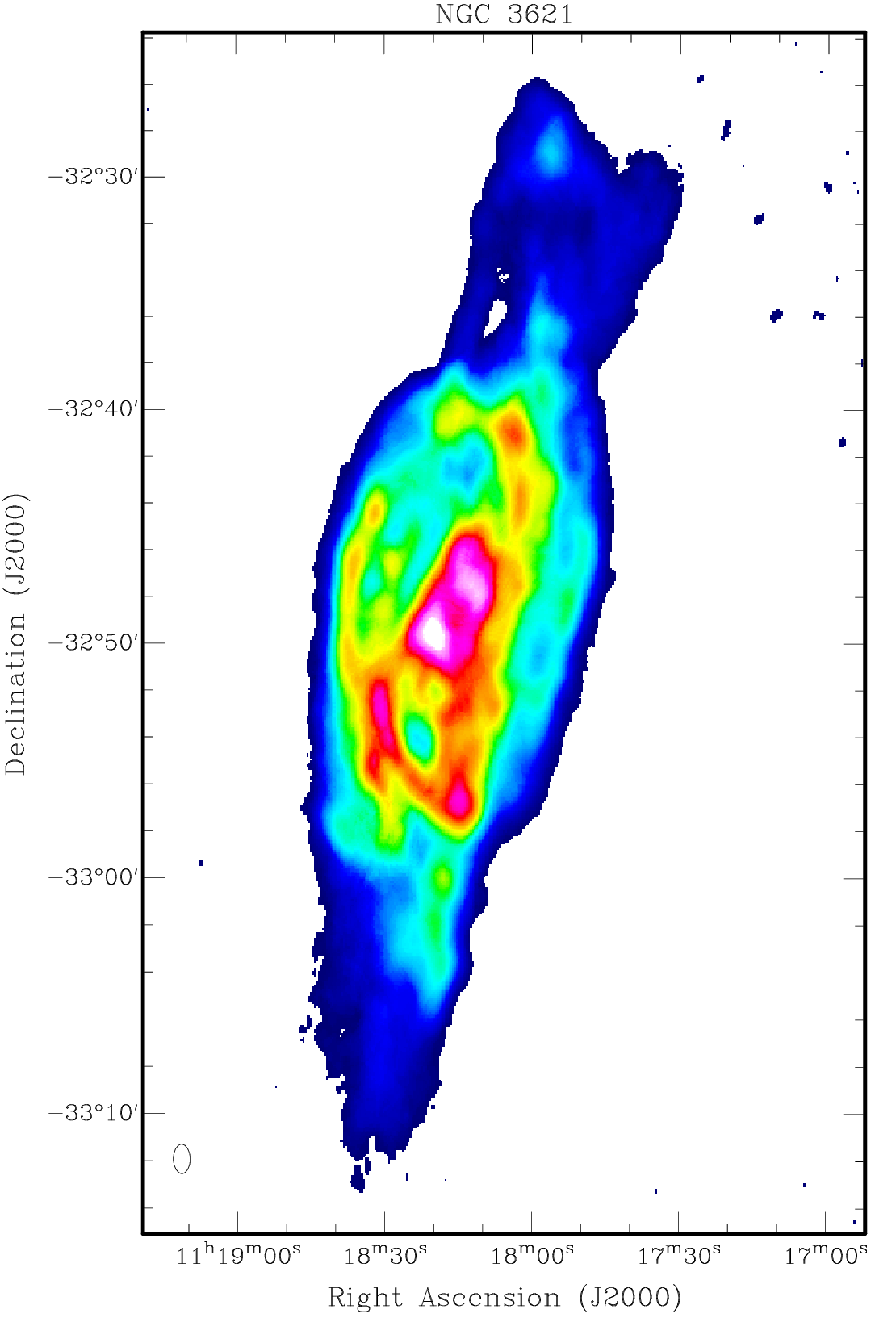} \\
\vspace*{-1.0cm}
\end{tabular}
\begin{tabular}{c}
 \includegraphics[width=9.8cm,angle=-90]{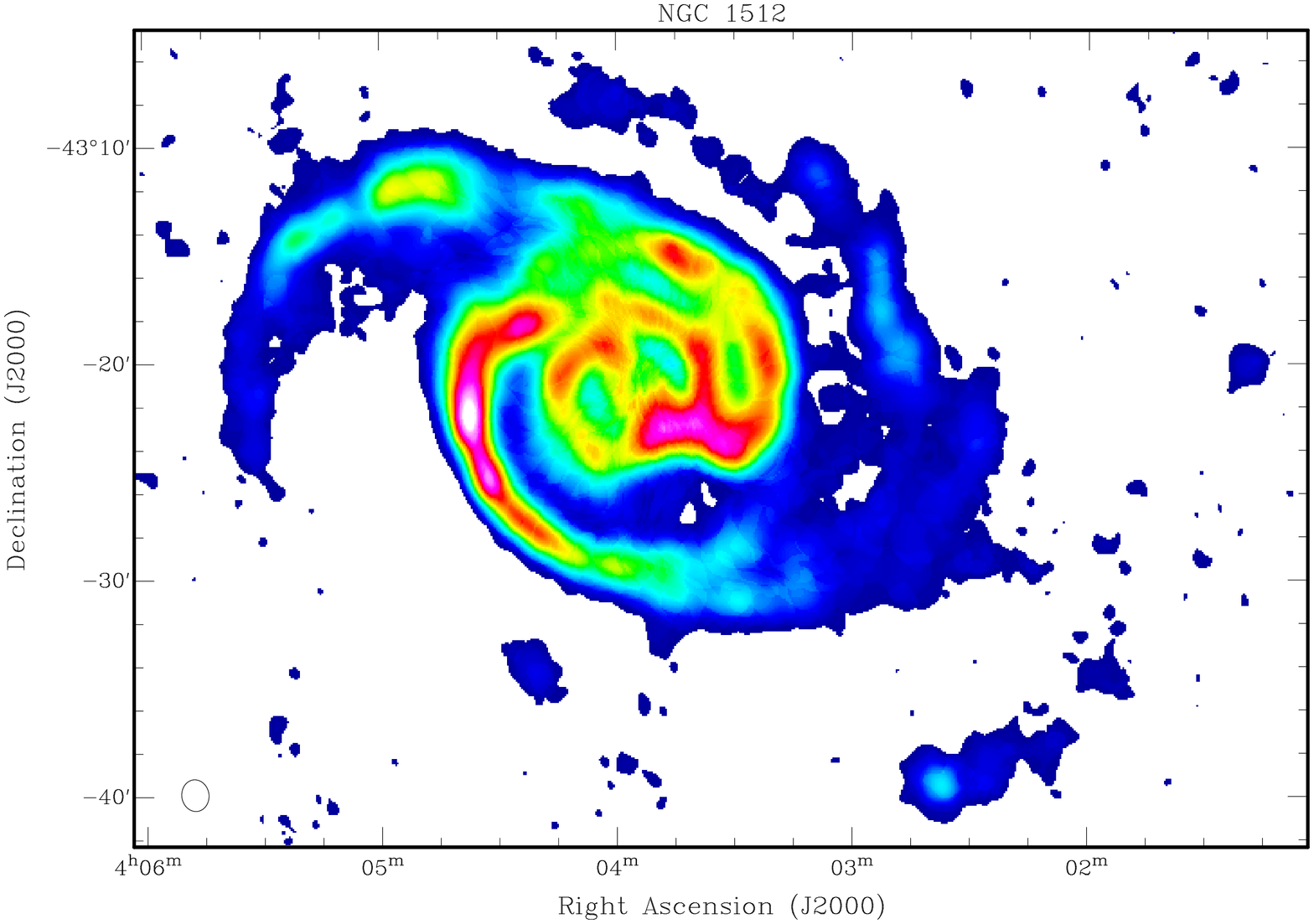} \\
\end{tabular}
\caption{2X-H\,{\sc i} disks of the spiral galaxies M\,83 (top left), 
   NGC~3621 (top right) and NGC~1512/1510 (bottom; \cite[Koribalski \&
   L\'opez-S\'anchez 2009]{KL2009}). The displayed ATCA H\,{\sc i} 
   mosaics are part of the Local Volume H\,{\sc i} Survey (LVHIS) 
   project (Koribalski et al. 2016). For more details see the project 
   webpage at {\em www.atnf.csiro.au/research/LVHIS}. }
\label{Koribalski.fig1}
\end{figure}

\section{The 2X-HI disks of nearby galaxies}

A search for more XUV disks by \cite[Thilker et al. (2007)]{Thilker2007}, 
revealed many more galaxies, among them the nearby, southern spirals NGC~300 
(\cite[Westmeier, Braun \& Koribalski 2011]{WBK2011}), NGC~1512 
(\cite[Koribalski \& L\'opez-S\'anchez 2009]{KL2009}), NGC~1672 
(\cite[Houghton 1998]{Houghton1998}), and NGC~3621 (\cite[Walsh 1997]{W1997}). 
ATCA H\,{\sc i} mosaics reveal their 2X-H\,{\sc i} disks, some of which are 
shown in Fig.\,\ref{Koribalski.fig1}. The galaxy H\,{\sc i} diameters are as
expected from their H\,{\sc i} masses and lie firmly on the $M_{\rm HI}$ --
$D_{\rm HI}$ relation (\cite[Wang et al. 2016]{Wang2016}; Wang, these 
proceedings). An H\,{\sc i} atlas of nearby galaxies is presented as part 
of the Local Volume H\,{\sc i} Survey (LVHIS; \cite[Koribalski 2008]{K2008}, 
\cite[Koribalski et al. 2016]{K2016}). LVHIS consists of a complete 
sample of $\sim$80 H\,{\sc i}-rich galaxies in the southern sky, observed with 
the Australia Telescope Compact Array (ATCA), $\sim$30 hours each, and 
supplemented by multi-wavelength images. 

The optical $B_{\rm 25}$ diameter of M\,83's bright stellar disk is $\sim$13$'$
(17~kpc at $D$ = 4.5 Mpc). Deep optical photography by \cite{MH1997}
reveals a much larger stellar disk ($\sim$20$'$) as well as faint streams 
and loops further out, indicative of dwarf galaxy accretion. The most prominent
stellar stream lies $\sim$20$'$ north of M\,83's center. For comparison, 
the H\,{\sc i} disk of M\,83, mosaiced by the ATCA (see 
Fig.\,\ref{Koribalski.fig1}), has a diameter of over 60$'$ (80 kpc), reveals a 
large tidal tail/arm wrapping at least 180 degrees around, from West to East,
and shows signs of ram pressure stripping on the compressed, western side. 
Also shown in Fig.\,\ref{Koribalski.fig1} are NGC~3621, which 
resembles M\,83 but seen at a larger inclination angle, and NGC~1512, 
which is known to interact with its blue compact dwarf companion NGC~1510.

GALEX provided a beautiful image of M\,83's XUV disk (\cite[Thilker et al. 
2005]{Thilker2005}), with UV emission detected out to a radius of $\sim$20$'$. 
The star-forming regions in M\,83's outer disk agree 
well with the high density H\,{\sc i} clumps, suggesting that H\,{\sc i} is 
an excellent tracer of star formation in the galaxy outskirts (see also 
\cite[Koribalski \& L\'opez-S\'anchez 2009]{KL2009}; 
\cite[For, Koribalski \& Jarrett 2012]{FKJ2012}).
A colorful overlay of M\,83's GALEX XUV-disk with the VLA single-pointing 
H\,{\sc i} map by \cite{Walter2008} featured on one of the IAU S321 conference 
posters, greeting participants every morning at our lovely venue in Toledo, 
Spain. We note that the H\,{\sc i} emission seen by the VLA in M\,83 and 
NGC~3621 (\cite[Walter et al. 2008]{Walter2008}) is only 22\% and 77\%, 
respectively, of the total H\,{\sc i} flux, which is well established from 
single-dish imaging (e.g., \cite[Huchtmeier \& Bohnenstengel 1981]{HB1981};
\cite[Koribalski et al. 2004]{K2004}).  \\

The ASKAP H\,{\sc i} All Sky Survey (WALLABY, DEC $<$ +30 degr) is expected to 
detect over 500\,000 galaxies and provide well-resolved maps for $\sim$5000 of 
these (\cite[Koribalski 2012]{K2012}).

\end{document}